# Investigation of charge states and multiferroicity in Fe-doped *h*-$YMnO_3$

Sonu Namdeo[1], A.K. Sinha[2], M.N. Singh[2], and A.M. Awasthi[1*]

[1]Thermodynamics Laboratory, UGC-DAE Consortium for Scientific Research,

University Campus, Khandwa Road, Indore- 452 001 (India).

[2]Indus Synchrotron Utilization Division, Raja Ramanna Centre for Advanced Technology,

Indore-452 013 (India).

## ABSTRACT

Polycrystalline $YMn_{1-x}Fe_xO_3$ ($YMFO_x$) ($0 \leq x \leq 0.1$) compounds have been prepared in single phase and characterized by synchrotron X-ray diffraction, X-ray absorption near edge spectroscopy, magnetization, and dielectric measurements. Iron-substitution in hexagonal $YMnO_3$ causes intra-lattice changes exceeding those of the lattice cell. XANES provide mixed-valence $Mn^{3+}/Mn^{4+}$ and $Fe^{4+}$ charge states in these manganites, consistent with the observed decrease of the effective magnetic moment with Fe-doping. Magnetization $M(T)$ evidence antiferromagnetic (AFM) ordering of the specimens with little weak ferromagnetism, and the metrices of exchange interaction suppress with Fe-doping, attributed to the lengthening of the Mn-O planar bond lengths. Dielectric $\varepsilon'(T)$ results showing highly doping-dependent anomaly at $T_N$ indicate linear magneto-electric coupling.

## I. INTRODUCTION

Multiferroics are a class of materials exhibiting both magnetic and ferroelectric orders in a certain temperature range.[1] They have promising potential in information-storage and spintronics such as non-volatile memory storage devices, ferro-electric random access memory (FERAM), actuators and magneto-electric (ME) sensors, and metal–ferroelectric–

---

[*] Corresponding Author. E-mail: amawasthi@csr.res.in. Tel.: +91 731 2463913. Fax: +91 731 2462294.

insulator–semiconductor (MFIS) field-effect-transistors.[2] A microscopic origin of multiferroicity has been enunciated in the hexagonal $YMnO_3$,[3-6] where the $MnO_5$ polyhedral blocks organize into trimers sharing a common planar oxygen O(3). Due to $MnO_5$ trigonal-bipyramidal environment (vis-à-vis that of $MnO_6$ octahedron in *o*-$RMnO_3$), double-occupancy of the crystal-field split low-lying $e_{1g,2g}$ orbitals removes the orbital degeneracy of the (unsplit $Mn^{+3}$-$3d^4$) ground state. No further energy gain from the otherwise degeneracy-lifting Jahn-Teller (J-T) lattice distortion therefore makes *h*-$YMnO_3$ and its Mn-substituents less J-T active.[7] The triangularly arranged Mn-spins,[8] interact (within the trimers) via the shared planar O(3,4) oxygens-mediated super-exchange. Intrinsic frustration (Mn-$x \approx 1/3$, large $f = |\theta_{CW}|/T_N$)[9] hinders their ordering until $T_N \sim$ 60-75K, much below their Curie-Weiss temperature $|\theta_{CW}| \sim$ 400-500K. Concomitant to the (A-type) AFM ordering, an isostructural transition causes a huge increase ($\geq$ 3.3%) in the atomic Mn-$x$ position,[6] comparable with that observed for the Ti atoms at the ferroelectric transition in $BaTiO_3$.[10] The super-super-exchange via the apical O(1,2) oxygens between the (AFM-ordered) Mn-spins of the adjacent basal-planes fixes their mutually antiparallel ($\Gamma_1$) configuration in Goldstone mode ($\boldsymbol{k} = 0$).[11] Magnetic frustration (trimerization) coupled tilting & buckling of $MnO_5$-bipyramids causing unequally-populated up/down displacements of Y-ions produces polarization, and ferroelectric order[4] below $T_C \sim$ 913K.[12] Spin-charge coupling in $YMnO_3$ has also been investigated in detail using Raman scattering.[13] Optical studies find a clear evidence for coupling between the ferro-electric (FE) and AFM domain walls.[14] Gigantic magneto-elastic (spin-lattice) coupling responsible for the magneto-electric (ME) effects in hexagonal manganites has been attributed to the interlocked FE domain walls and structural antiphase boundaries, latter becoming the AFM boundaries below $T_N$.[14-16] At the atomic level, (Hamiltonian) configuring the in-plane spin-order and out-of-plane (vertical) non-centrosymmetry is yet to be established.

In studies on B-site substitution mostly by the non-magnetic dopants, a direct change of the lattice constants (due to the guests' ionic radii being rather different from Mn's) dominantly affects the biferroic electrical[17] and magnetic[18] orders in $YMnO_3$; obscuring their coupling aspect, which is minimally influenced.[19-20] On the other hand, a magnetic dopant (especially Fe, with $r_{Fe} \approx r_{Mn}$) ought to keep the basic lattice vectors rather intact and instead alter the magneto-strictive (MS) intra-lattice features exclusively, which can be determined to high accuracy by synchrotron diffraction data, unhindered by any other (thermo-mechanical) baseline dominances, as our study reveals. So far, only a few isolated investigations have focused on the Fe-doping in hexagonal $YMnO_3$, mostly concerned with the doping limit,[21-22] before the structure transforms to orthorhombic. Also, the exact order of magneto-electric coupling here remains experimentally unestablished, or is only alluded to.[5,14]

Moreover, only a few works on $YMnO_3$ have provided a consistent fixation of the charge/valance state of the B-cations.[21-22] Generally, iron replacing manganese as $Fe^{+2}$ ($Fe^{+3}$) decreases (increases) the magnetic moment of the system; the reporting by Zaghrioui et al[22] of the $Fe^{+3}$ charge state in YMFO by Mössbauer spectroscopy though contradicts the observed decrease of the effective magnetic moment. Partial replacement of Mn may also change the average charge state of the remaining Mn; thereby additionally changing the magnetic moment. Therefore, reliable fixation of the actual charge states of Mn and Fe is of essence. To this end, we have carried out X-ray absorption near edge structure (XANES) spectroscopy, which is a powerful element-specific tool, extremely sensitive to the electronic states as well as the local co-ordination and hybridization effects of the elements present in the sample.[23] In literature, there is no report of XANES study on YMFO, to the best of our knowledge. We address the above important issues, by studying the effects of perturbative chemical substitution, well within the saturation-changes and without major structural modifications. Here we present results of our investigations spanning a wider set of

experiments--- synchrotron-based diffraction & absorption, together with the magnetic and dielectric characterizations. Our analysis indicates a linear magneto-electric coupling which far exceeds that observed with non-magnetic Mn-site substitution.

## II EXPERIMENTAL

Polycrystalline samples of $YMFO_{x=0-0.1}$ were synthesized by conventional solid state reaction route. Stoichiometric amounts of high purity (>4N) $Y_2O_3$, $MnO_2$, and $Fe_2O_3$ powders were thoroughly mixed and calcined at 1200 ºC for 12 hrs. The resultant powders were pelletized and sintered in air at 1400 ºC for a total of 12 hrs, with intermediate grindings. To check the purity of the specimens, the as-prepared samples were subjected to X-ray diffraction (Rigaku with Cu-$K_\alpha$). All the specimens were found single-phase, without any detectable impurities. Microstructure and chemical composition were analyzed with Scanning Electron Micrometer (JEOL JSM 5600), equipped with an Oxford EDS Spectrometer (INCA). Synchrotron X-ray powder diffraction (SXRD) and X-ray absorption near edge structure (XANES) were carried out at angle-dispersive X-ray diffraction (ADXRD) beamline (BL-12) on Indus-2 synchrotron source.[24] The beamline has two experimental stations, namely a six-circle diffractometer with a scintillation-point-detector and an **Mar 345** image-plate area detector. SXRD was carried out using the image-plate. The X-ray wavelength used for the present study was accurately calibrated by doing XRD on the $LaB_6$ NIST standard. Data reduction was done using Fit2D software.[25] X-ray absorption near edge structure (XANES) was carried out at the same beamline. The high-resolving-power beamline consists of a Si (311) crystal-pair based double-crystal monochromator. The beam-energy bandpass at 8keV is less than 0.8 eV, and its repeatability is ~50meV. The Mn- and Fe- K-edge spectra were collected for YMO and $YMFO_{x=0.05-0.1}$ samples respectively, along with the same for $Mn_3O_4$, $Mn_2O_3$, $MnO_2$, FeO, $Fe_3O_4$, and $Fe_2O_3$ as the reference compounds. The measurements were carried out in fluorescence mode using an energy dispersive detector (VERTEX-EX). The spectra were normalized to the incident photon current, measured by an ionization chamber. Magnetization data were taken using a Quantum Design-14T magnetic properties measurement system (MPMS)-vibrating sample magnetometer (VSM).[26] For carrying out the dielectric measurements, powder-samples were pelletized into 10 mm diameter tablets and their planar surfaces were polished flat. For proper electrical contact, high-temperature silver-paint was applied onto their flat surfaces. Pellets were then dried and fired at 500 °C for 15 min, to stabilize the applied paint. The measurements were performed using Novocontrol Alpha-A High Frequency Analyzer, in concert with a homemade test cell (parallel-plate capacitor configuration), scanned across 10Hz-10MHz over room temperature down to 10K range.[26]

## III RESULTS AND DISCUSSION

The samples' morphology were found to be uniform, with average grain size on the order of ~8μm. Energy dispersive analysis of x-ray (EDAX) results ensure that the samples are chemically homogeneous, with the ratio of site-cations being close to the nominal value. Precise details of structural distortion (lattice constants, bond lengths/angles etc) were determined from the SXRD by performing structure-refinement using FULLPROF-2K software. Fig.1a shows the Rietveld-calculated diffraction pattern of $YMFO_{x=0.05}$ specimen as a representative one. The difference between the observed and calculated patterns is appreciably small with the goodness of fit coming out as 1.4. The same fitting strategy was followed for all the compositions. The refined lattice constants were found to be $a = b$ =6.1471(5)Å & $c$ =11.3704(1)Å for $YMFO_{x=0,}$ $a = b$ = 6.1487(2)Å & $c$ =11.3715(3)Å for $YMFO_{x=0.05,}$ $a = b$ = 6.1594(2)Å & $c$ =11.3892(2)Å $YMFO_{x=0.08}$, and $a = b$ = 6.1489(6)Å & $c$ =11.3771(2)Å for $YMFO_{x=0.1}$. Similar increasing trend of the lattice constants were reported by Zaghriouri et.al.[22] However, the nanocrystalline YMFO reported by Han et.al.[27] showed the opposite trend ($1/a$ linear in $x$). The difference between lattice constants obtained in specimens prepared via bulk- and nano-synthesis routes reflects the larger oxygen non-stoichiometry (surface to volume ratio) in the latter, which gets enhanced with the chemical disorder (e.g., the dopant level). As the dopant content ($x$) is increased, the differential oxygen non-stoichiometry rises, so much so that one observes opposite trends of $a(x)$ in the specimens thus differently synthesized. To further analyze, we determine in-plane bond-lengths and bond-angle from the refined structure data (Table I). The increasing average planar bond-length and the decreasing planar bond-angle reflect reducing planar exchange interaction, suggesting the suppression of $T_N$ and were found to be lowest in YMFO$_{x=0.08}$ sample, which will be further confirmed by magnetic measurements. The analogous $YbMn_{1-x}Fe_xO_3$ ($0 \leq x \leq 0.3$) system shows opposite variations of these bond-lengths/angles, as well as the change in the Néel temperature $T_N$.[28] This can be understood as follows. Mn-ions form a nearly ideal two-dimensional triangular lattice,[17] with two Mn-O planes per unit cell. Across the critical Mn-$x$ =1/3 position, the inter-plane Mn-Mn super-super-exchange interaction via apical oxygens (O1/O2) changes its sign; the $\Gamma_1$ ($\Gamma_4$) configuration corresponds to Mn-$x$ >1/3 (<1/3), where Mn moments are perpendicular to $a$- and $b$- axes, and their arrangement in the $z$ = 1/2 plane is antiparallel (parallel) with respect to those in the $z$ =0 plane.[11] Different arrangements ($\Gamma_1$ vs. $\Gamma_4$) realized therefore cause opposite changes in isostructural & magnetic parameters of (Y vs. Yb)$Mn_{1-x}Fe_xO_3$, upon Fe-doping.

In order to pin the charge state of Mn and Fe in YMFO$_{x=0-0.1}$, we have carried out X-ray absorption near edge structure (XANES) spectroscopy at Mn and Fe K-edges, respectively. Figurer 2 shows the normalized Mn K-edge XANES spectra of samples, along with three reference samples of manganese oxide with different charge state of Mn i.e., $Mn_3O_4$ (+2.6), $Mn_2O_3$ (+3), and $MnO_2$ (+4). The feature marked 'P' (the pre-edge) and 'E' (the edge) in the figure are due to the 1s-3d and 1s-4s quadruple-allowed transitions respectively, while the feature marked as 'W' is known as the white line, and corresponds to the dipole 1s-4p transitions.[29-30] In addition, one can see a prominent feature marked as 'B' appearing beyond the while line in both Mn and Fe K-edge spectra (figs.2 and 3). We believe that this common feature may be due to the partial density of states of Y (yttrium). It is evident from fig.2 that the main edge of samples lies between $Mn_2O_3$ (+3) and $MnO_2$ (+4). It is well known that the main absorption edge shifts towards the higher energy side for higher charge/valance state.[31] The absorption edge energy is usually defined by the first inflexion point (taken as the absorption-coefficient value of ~ 0.8), and is often used for comparisons of charge states. To further elucidate the chemical shifts, we have tabulated the value of energy $E_0$ for reference manganese oxides and $YMFO_{x=0-0.1}$ samples in Table-II. The energies were found to be ($E_0$ ~ 6554eV) for our YMFO samples, which lie between those of $Mn_2O_3$ ($E_0$ ~ 6553eV) and $MnO_2$ ($E_0$ ~ 6556eV). This clearly vindicates that $YMFO_{x=0-0.1}$ samples have a mixed state of $Mn^{+3}$ and $Mn^{+4}$. Also, as we increase the Fe-content, there is a very little shift of the edge towards +4 charge state in fig.2.

Figure 3 shows normalized Fe K-edge XANES spectra of Fe-doped $YMFO_{x=0.05-0.1}$, with three standards of iron oxide i.e., FeO (+2), $Fe_3O_4$ (+2.6), and $Fe_2O_3$ (+3). As one can see the edges of $YMFO_{x=0.05-0.1}$ samples lie above all the reference samples and occur on higher energy side. This confirms that like Mn, Fe has a charge state of more than +3. We have also tabulated the energy $E_0$ for $YMFO_{x=0.05-0.1}$ specimens, from Fe K-edge data (Table-III). The K-edge energy for +4 charge state of Fe in $SrFeO_{3-\delta}$ has been reported[32] as ~7128eV. We find the energy $E_0$= 7128 ± 0.5eV for our $YMFO_{x=0.05-0.1}$, clearly inferring the presence of +4 Fe charge state. Moreover, it is interesting to see that Mn and Fe K-edge XANES spectra are quite similar. This further confirms that Fe is replacing Mn-site only. In addition, $Fe^{+4}$ can replace $Mn^{+3}$ or $Mn^{+4}$. If $Mn^{+4}$ gets replaced then there will be increase in the magnetic moment of the system and reduction of the average Mn charge state, without any charge imbalance, contrary to our observations. However, if $Fe^{+4}$ replaced $Mn^{+3}$, there will be no change in the magnetic moment as such but charge imbalance will ensue. In the latter case, some of the remaining $Mn^{+3}$ may convert to $Mn^{+4}$, thereby decreasing the magnetic moment (as observed from the analysis of our magnetization data in the following) and increasing the average Mn charge state (as also indicated in our XANES results).

To study the doping effects on the magnetic properties, we measured DC magnetization over 10-300K in an applied field of 0.1T shown in fig.4, with little different zero-field cooled (ZFC) and field-cooled (FC) data. The samples exhibit magnetic ordering as small kinks; their transition temperatures were more obvious from the differentiated curves (*dM*/*dT*, ZFC). $T_N$ obtained from the discontinuity in the slope was found to decrease minutely (uncertainty ±0.1K, Table IV), consistent with planar bond-lengths' increase with Fe-doping (largest in $YMFO_{x=0.08}$). Little difference in FC and ZFC magnetizations implies no weak ferromagnetism (WFM). The magnitude/decrease of $T_N$ fairly matches with a recent report on YMFO, prepared via the low temperature chemical route,[22] but which, contrary to our results, shows a pronounced (Fe-content-dependent) splitting in FC and ZFC data below the AFM transition. Also our magnetization at the lowest temperature (~10K) decreases with Fe-doping, again with the exception of $YMFO_{x=0.08}$.

We evaluated the Curie-Weiss temperature ($\theta_{CW}$) and estimated effective magnetic moment ($\mu_{eff}$) from the linear part of the inverse-susceptibility at high temperatures (fig.4, lower inset), and the results are summarized in Table IV. These parameters signifying magnetic correlations above $T_N$ and determining $Mn^{+3}/Mn^{+4}$ admixture respectively exhibit appreciable changes with Fe-doping. High values of $|\theta_{CW}|$ vis-à-vis $T_N$ manifest in the heat capacity as releasing a large part of magnetic entropy above $T_N$.[6] Lowering of $|\theta_{CW}|$ with Fe-doping therefore piles up a larger magnetic-entropy change at $T_N$. This, combined with the enhanced $M(T_N)$ give more prominent AFM transition in the doped specimens, as also observed in $M(T)$. The decreasing values of $\mu_{eff}$ too signify the $Fe^{+4}$ state. The Mn and Fe K-edge XANES confirm mixed valance state of Mn ($Mn^{+3}/Mn^{+4}$), and iron in +4 charge state. Also, the magnetic moments ($\mu_{eff}$) of $Mn^{+3}(3d^4)$, $Mn^{+4}(3d^3)$, and $Fe^{+4}$ $(3d^4)$ are respectively, $4.9\mu_B$, $3.87\mu_B$, and $4.9\mu_B$. This leads to the conclusion that with increasing Fe-content, magnetic moment is expected to be suppressed, as per our observations.

The ratio $f \sim (|\theta_{CW}|/T_N)$ as the metric of frustration in the magnetic structure[33] decreases upon Fe-doping (Table IV, with consistent exception of $YMFO_{x=0.08}$). In fig.4 (upper inset), magnetic frustration $f$ and the effective magnetic moment $\mu_{eff}$ are seen as covariant, camouflaging any compositional excursions. Decrease of $\mu_{eff}$ with iron-doping was earlier argued to be consequent to the $Fe^{+4}$ charge state of the doped iron, causing the mixed-valence $Mn^{+3}/Mn^{+4}$ (charge-disorder). Since the same Mn-ions that bear the moments ($\mu_{eff}$) undergo this charge-disorder, the latter by lifting magnetic degeneracy breaches their spin-frustration $f$ (Mn-$x \rightarrow > 1/3$); the allied untilting/unbuckling of $MnO_5$-polyhedra tends to restore the out-of-plane centrosymmetry in the YMFO system, as also evident from the high-temperature powder neutron diffraction data showing complete demise of the MnO5 polyhedral buckling above $T_C$.[34]

The dielectric constant $\varepsilon'$(10kHz) measured for $YMFO_{x=0-0.1}$ specimens over temperature range 30K-300K is shown in fig.5. Here, a rapid rise observed in $\varepsilon'(T)$ above $T$~250K is probably due to the electron hopping between $Mn^{+3}$ and $Mn^{+4}$, traced to oxygen deficiency.[35] These results are also in agreement with our Mn-K edge XANES results (fig.2), which confirms the presence of mixed valance state of $Mn^{+3}/Mn^{+4}$. Left inset depicts the close-up $\varepsilon'(T)$ behavior, with clear signature-drops near $T_N(x)$ in all four $YMFO_{x=0-0.1}$, confirming the presence of magneto-electric coupling in pure[36] as well as in Fe-doped specimens. Suppression of the $\varepsilon'(T_N)$-anomaly at higher Fe-content is consequent to its manifestation at a higher polarization $P_z$ (as $T_N(x) < T_N(0)$).[5] Simultaneous smearing of the $\varepsilon'(T_N)$-anomaly with doping relates to the lower slope of the mean-field curve at $T_N(x) < T_N(0)$.

To quantify the magneto-electric (ME) effect, we have measured the relative shift at $T_N(x)$ in the dielectric constant, evaluated as $(\Delta\varepsilon/\varepsilon_{HT})_{T_N}$, and found that, our 'magneto-electric metric' i. e., $(\Delta\varepsilon/\varepsilon_{HT})_{T_N}$ and its Fe-content-dependence (-10dB) far exceed a decade-smaller magneto-capacitance reported by Aikawa et. al.[20] in $YMn_{1-x}Ti_xO_3$. This affirms our assertion that the magnetic (non-magnetic) Mn-substituents mostly altering the MS-intra-lattice (lattice) attributes essentially affect the mutual-coupling (individual-features) of the biferroic orders. Furthermore our metric of ME effect $(\Delta\varepsilon/\varepsilon_{HT})_{T_N}$ scales linearly (right inset) with $(\Delta M/M_{CW})_{T_N}$; latter evaluated by the difference in the measured and Curie-Weiss-extrapolated magnetizations at $T_N$ (inset). From ($P = \varepsilon E$) we get $(\Delta P/P)_{T_N} = -(\Delta\varepsilon/\varepsilon)_{T_N}$ (variations of spontaneous (dc) polarization and measured (ac) polarizability/permittivity being opposite in sign). The linearity thus indicated between the relative changes (vs. doping) of polarization and magnetization at $T_N$ could originate only from their homogeneous-coupling term, e.g. $-\gamma(PM)^{\delta}$, in a free-energy description; invoking a correspondence between Fe-doping and sub-$T_N$-cooling. Lee et. al.[5] have argued for thus-coupled spin/lattice and lattice/dipole orders within the Ginzburg-Landau framework, with a common mean-field character displayed by suitably normalized polarization ($\Delta P_z$), effective-moment ($\mu_{ord}$), and planar-bond-split $|d_{Mn\text{-}O(3)}\text{-}d_{Mn\text{-}O(4)}|$. These universally-scaled (overlapping) parameters shown by them as representing the electrical and magnetic orders suggest their mutually linear covariance in the ordered state. Variation vs. the Fe-content ($x$) in $\Delta\varepsilon(x)|_{T_N}$ (just as in $\Delta M(x)|_{T_N} = \{M - M_{CW}\}_{T_N}$) originates from the changes in spontaneous internal-field. We propose that the observed linearity between these two allied differentials corroborates the linear magneto-electric character of the system, although a rigorous confirmation requires measurements of permittivity under applied magnetic field. Fiebig et. al.[14] suggest a linear coupling (*Pl*) of electrical (FE) and magnetic (AFM) order-parameters at the domain-walls; the bulk-level linear coupling forbidden by symmetry in $YMnO_3$.[37]

## II. CONCLUSIONS

Despite the anomalous composition- dependences, low-level Fe-substitutions on Mn-site in $YMnO_3$ multiferroic, clearly contrast the roles of magnetic versus non-magnetic B-site-dopants. Synchrotron XRD and XANES provided precise sublattice parameters and fixated the exact +4 charge state of the Fe-dopant, with emergent insights. The doped magnetic Fe ions weaken the Mn-Mn interaction which results in a decreased value of effective paramagnetic moment ($\mu_{eff}$) and Curie-Weiss temperature ($\theta_{CW}$). Doping-driven enhanced charge-disorder ($Mn^{+3}/Mn^{+4}$) quenches the spin-frustration (*f*); this affects the biferroicity minutely, but alters their coupled magneto- electrical features hugely. A linear regression between the relevant magnetic and electrical susceptibility-components at $T_N$ is recognized.

## ACKNOWLEDGEMENTS

We thank D.M. Phase for SEM/EDAX micrographs and A. Banerjee for magnetization data. S.K. Deb (Synchrotron Utilization Division, RRCAT, Indore) is thankfully acknowledged for the XANES work. P. Chaddah and A. Gupta are acknowledged for their scientific support.

## Table Titles

**Table I.** Variation of In-plane bond-length (Å) and bond-angle (deg) with the Fe-content.

**Table II.** Energy $E_0$ for reference samples $Mn_3O_4$, $Mn_2O_3$, $MnO_2$, and $YMFO_{x=0\text{-}0.1}$.

**Table III.** Energy $E_0$ for reference samples FeO, $Fe_3O_4$, $Fe_2O_3$, and $YMFO_{x=0.05\text{-}0.1}$.

**Table IV.** Variation of AFM transition temperature $T_N$ (to within ±0.1K), Curie-Weiss temperature $\theta_{CW}$, effective magnetic moment $\mu_{eff}$, and frustration parameter $f$.

## Figure Captions

**Figure 1.** Rietveld-refined SXRD pattern for $YMFO_{x=0.05}$.

**Figure 2.** Normalized Mn-K edge XANES spectra of $Mn_3O_4$ (+2.6), $Mn_2O_3$ (+3), $MnO_2$ (+4), and $YMFO_{x=0\text{-}0.1}$ samples. Here 'P', 'E', and 'W' stand respectively for the pre-edge, edge, and white-line features, while 'B' (commonly appearing with Fe-K edge XANES) most probably corresponds to the Y-partial density of states. Spectra vertically offset for visibility.

**Figure 3.** Normalized Fe-K edge XANES spectra of FeO (+2), $Fe_3O_4$ (+2.6), $Fe_2O_3$ (+3) and $YMFO_{x=0.05\text{-}0.1}$ samples. For the sake of visibility the spectra have been offset vertically.

**Figure 4.** Measured *M-T* for $YMFO_{x=0\text{-}0.1}$ display kinks at the Néel temperature $T_N(x)$. Lower inset: $1/\chi$ (=$H/M$) with Curie-Weiss linear *T*-dependence near the room temperature. Upper inset: anti-regression of the metrices of spin-frustration ($f = |\theta_{CW}|/T_N$) and effective magnetic moment $\mu_{eff}$.

**Figure 5.** $\varepsilon'$-$T$ plot for $YMFO_{x=0\text{-}0.1}$ at 10 kHz (similar at all $\omega$'s; curves are *y*-shifted for clarity). Left inset: zoomed-in magneto-electric (ME) anomaly at $T_N$ and its rapid (logarithmic) reduction with the Fe-content. Right inset showing $(\Delta\varepsilon/\varepsilon_{HT})_{T_N}$ vs. $(\Delta M/M_{CW})_{T_N}$ implies a linear ME coupling (symbol-size represents the Fe-content).

## Table I

| Atoms | $YMFO_{x=0}$ | $YMFO_{x=0.05}$ | $YMFO_{x=0.08}$ | $YMFO_{x=0.1}$ |
|---|---|---|---|---|
| **In-plane bond-length** | 2.0615(6) | 2.0693(5) | 2.0718(5) | 2.0706(4) |
| **In-plane bond-angle** | 115.4(4) | 114.6(4) | 115.2(3) | 114.5(5) |

## Table II

| Sample | Energy $E_0$ (eV) |
|---|---|
| $Mn_3O_4$ | ~ 6552.94 |
| $Mn_2O_3$ | ~ 6553.28 |
| $YMFO_{x=0}$ | ~ 6554.04 |
| $YMFO_{x=0.05}$ | ~ 6554.10 |
| $YMFO_{x=0.08}$ | ~ 6554.12 |
| $YMFO_{x=0.1}$ | ~ 6554.19 |
| $MnO_2$ | ~ 6556.62 |

## Table III

| Sample | Energy $E_0$ (eV) |
|---|---|
| FeO | ~ 7122.68 |
| $Fe_3O_4$ | ~ 7124.87 |
| $Fe_2O_3$ | ~ 7125.54 |
| $YMFO_{x=0.05}$ | ~ 7128.03 |
| $YMFO_{x=0.08}$ | ~ 7128.25 |
| $YMFO_{x=0.10}$ | ~ 7128.26 |

## Table IV

| Composition | $T_N$ (K) | $\|\theta_{CW}\|$ (K) | $\mu_{eff}$ ($\mu_B$) | $f$ ($\|\theta_{CW}\|/T_N$) |
|---|---|---|---|---|
| $YMFO_{x=0}$ | 68 | 437 | 5.3 | 6.43 |
| $YMFO_{x=0.05}$ | 67 | 419 | 5.2 | 6.25 |
| $YMFO_{x=0.08}$ | 66 | 312 | 4.7 | 4.72 |
| $YMFO_{x=0.1}$ | 67 | 382 | 4.95 | 5.70 |

**Fig.1**

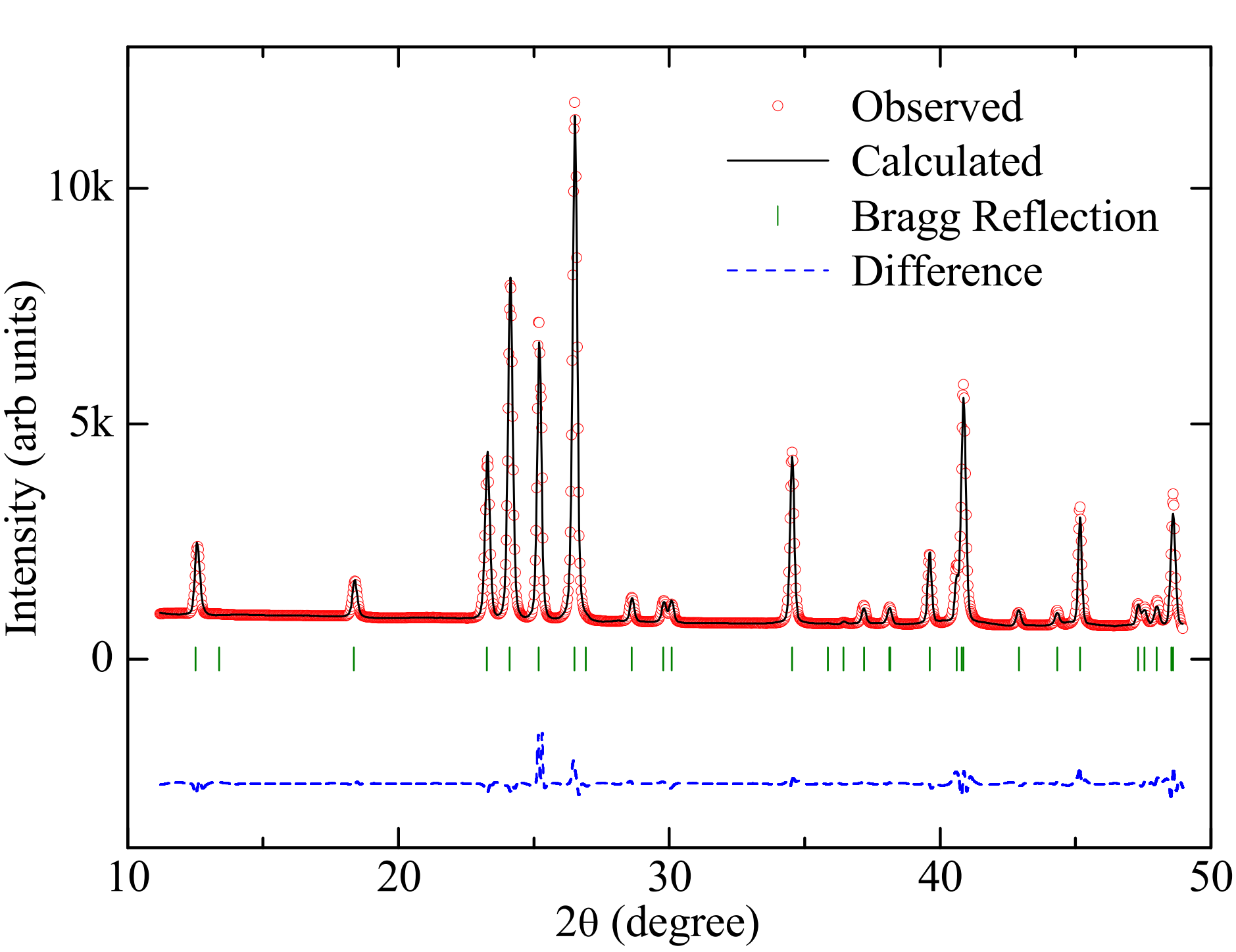

**Fig.2**

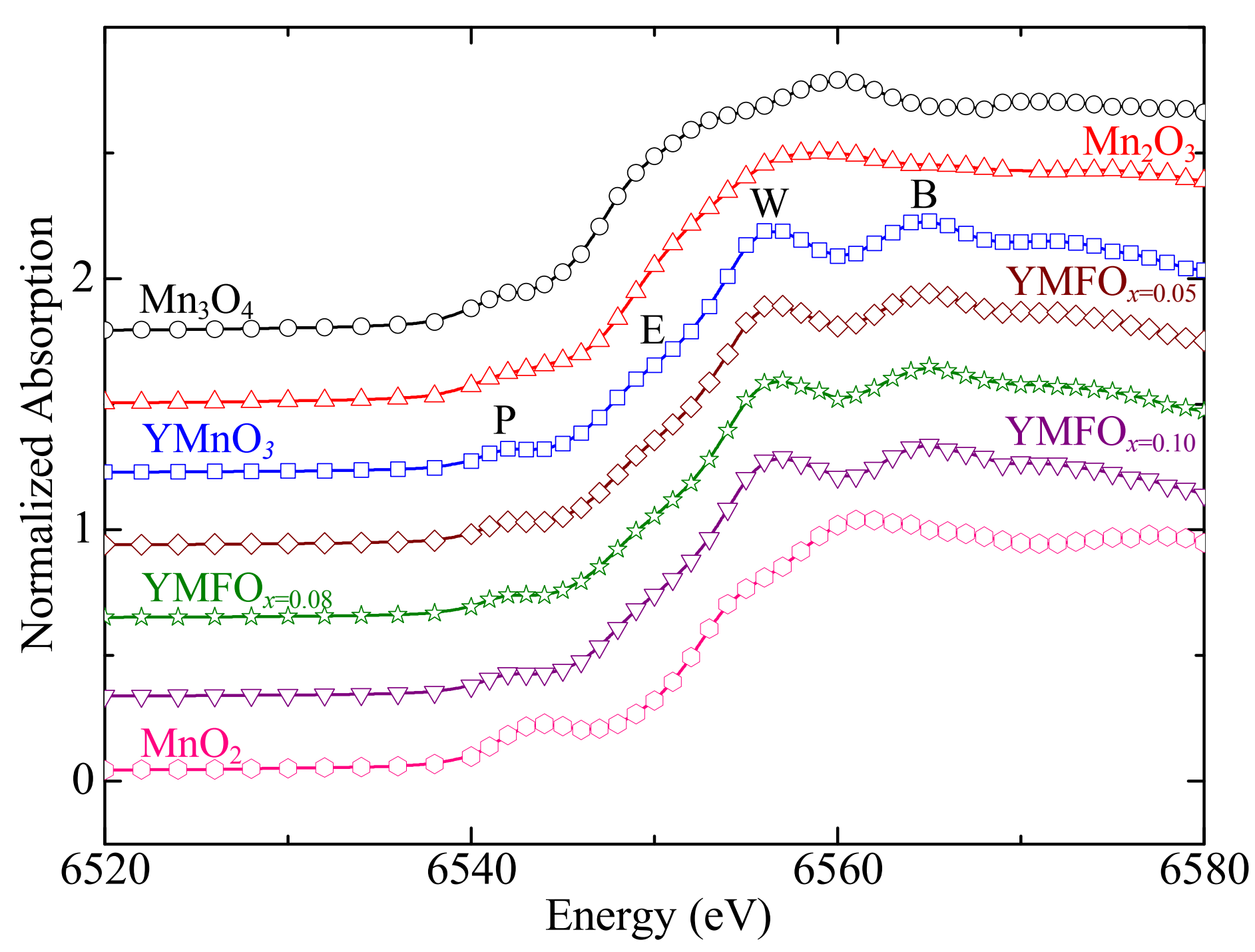

**Fig.3**

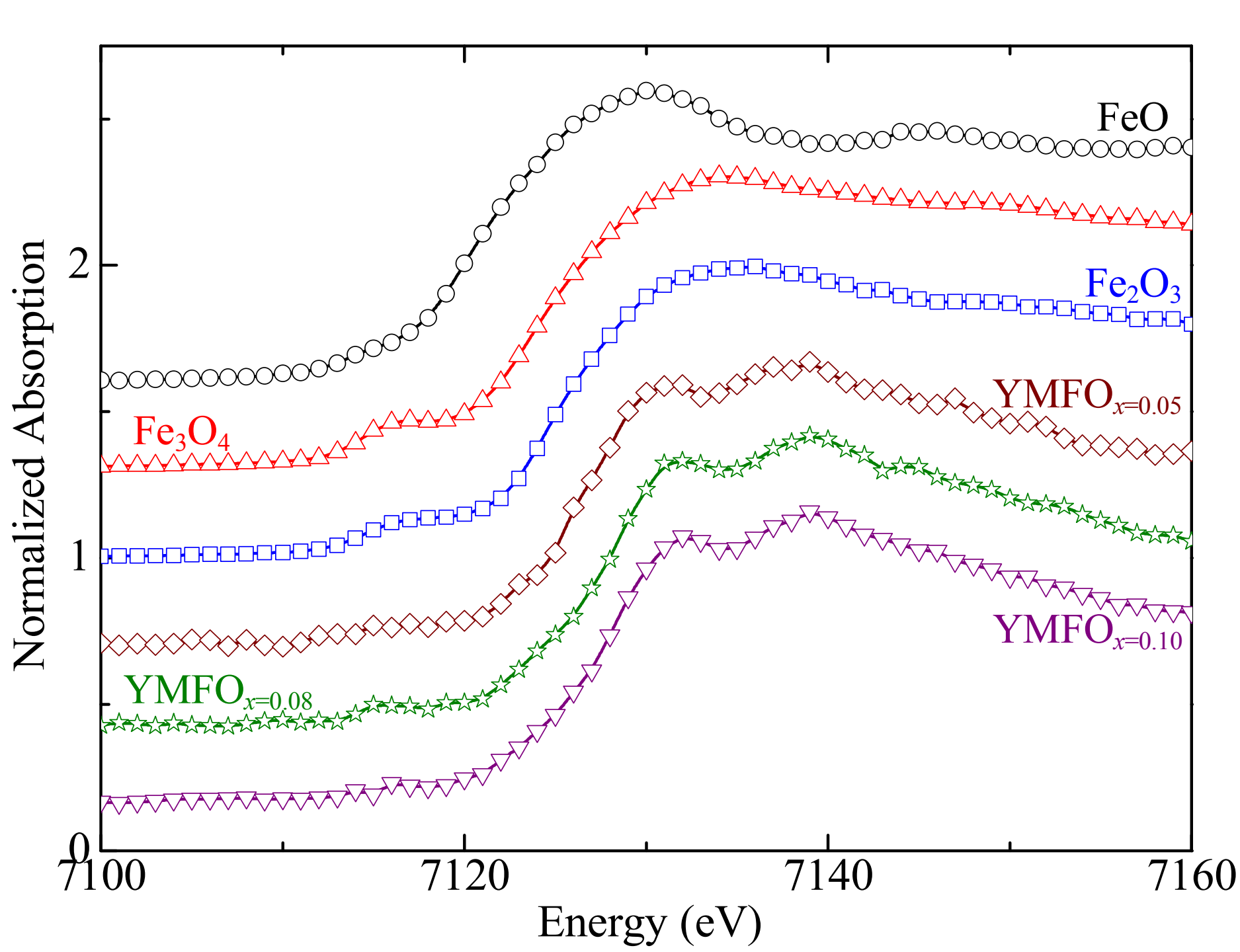

**Fig.4**

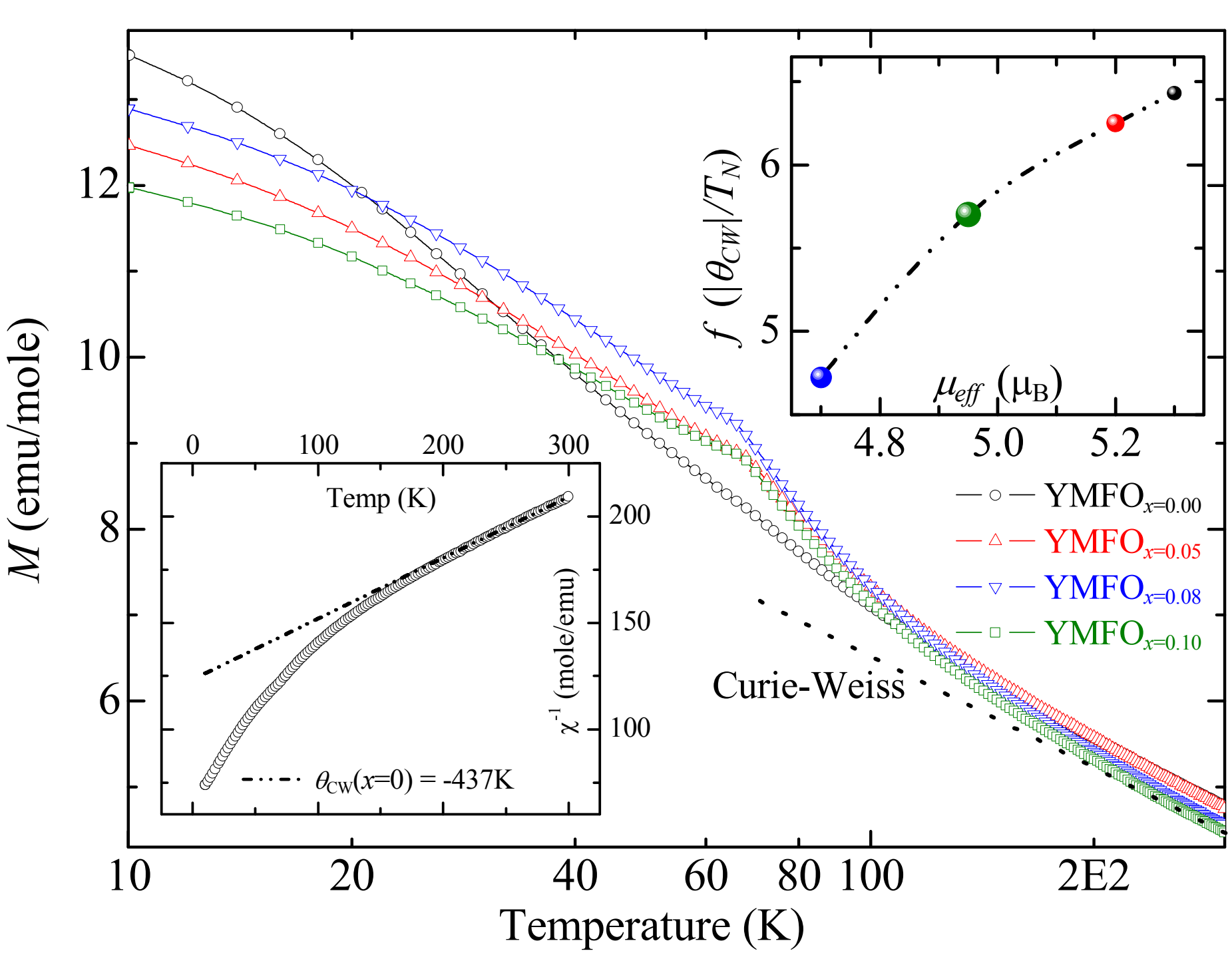

**Fig.5**

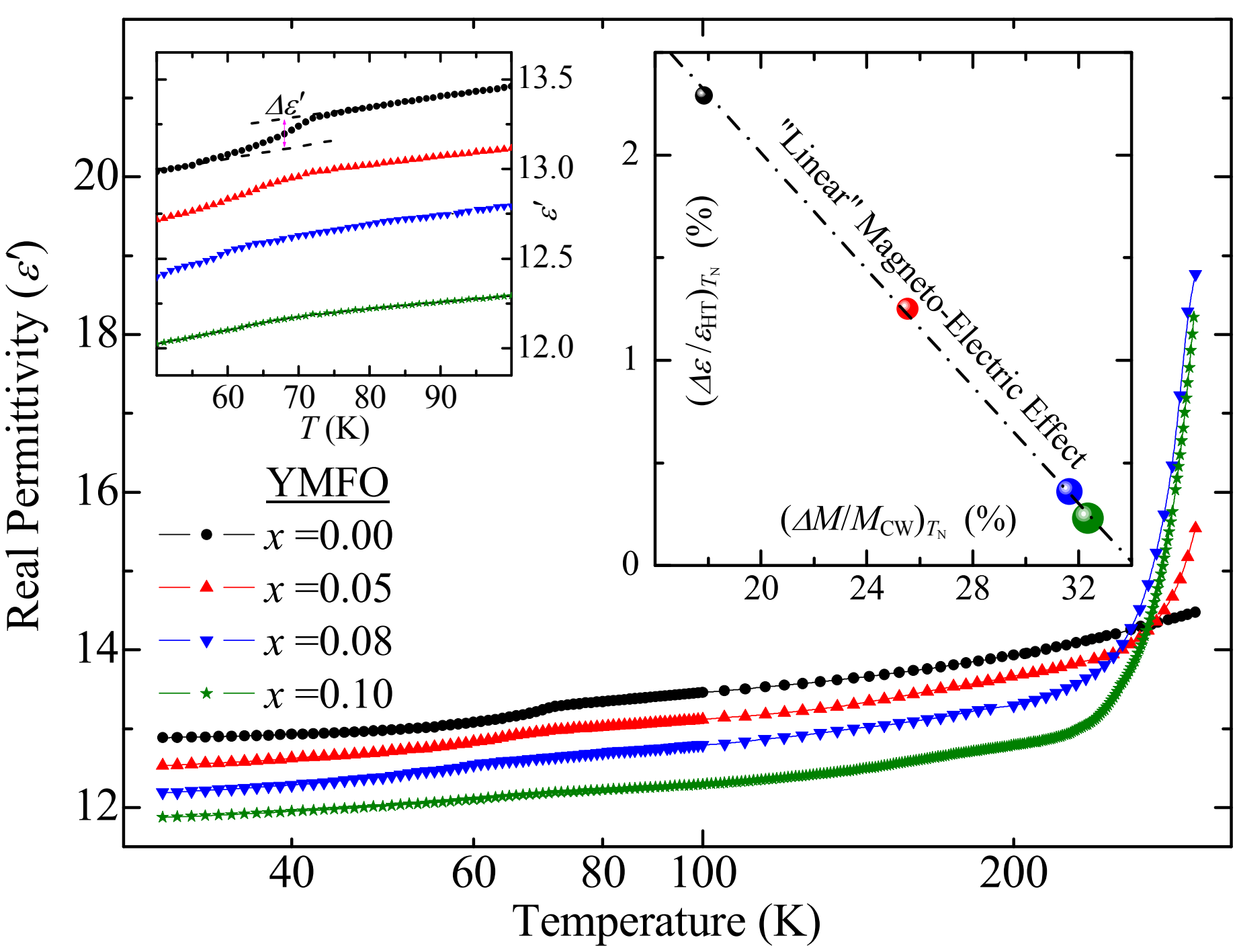